\documentclass[amsmath,amssymb,aps,twocolumn]{revtex4}
\usepackage{amsthm,amsfonts,graphicx,verbatim}
\usepackage{amsmath}
\usepackage{amssymb}
\usepackage{amsthm}
\usepackage{amsfonts}
\usepackage{listings}
\usepackage{enumerate}
\usepackage{latexsym}
\usepackage{psfrag}
\usepackage{bm}
\usepackage{graphicx}
\usepackage{subfigure}
\usepackage{float}

\usepackage{color}

\newcommand{\ket}[1]{\left| #1 \right>} % for Dirac kets
\newcommand {\ii} {i}

\begin{document}

\title{Experimental realization of strong effective magnetic fields \\ 
in optical superlattice potentials}

\author{Monika Aidelsburger,$^{1,2}$ Marcos Atala,$^{1,2}$ Sylvain Nascimb\`ene,$^{1,2,*}$\\ Stefan Trotzky,$^{1,2,\dagger}$ Yu-Ao Chen,$^{1,2,\ddagger}$ and Immanuel Bloch$^{1,2}$ }
\affiliation{%
$^{1}$Fakult\"at f\"ur Physik, Ludwig-Maximilians-Universit\"at, Schellingstr. 4, 80799 Munich, Germany\\
$^{2}$Max Planck Institute of Quantum Optics, Hans-Kopfermann Str. 1, 85748 Garching, Germany\\
$^{*}$Present address: Laboratoire Kastler Brossel, CNRS, UPMC, Ecole Normale Sup\'erieure, 24 rue Lhomond, 75005 Paris, France\\
$^{\dagger}$Present address: Department of Physics, University of Toronto, Toronto, Ontario, Canada M5S 1A7\\
$^{\ddagger}$Present address: Shanghai Branch, National Laboratory for Physical Sciences at Microscale and Department of Modern Physics, University of Science and Technology of China, 201315 Shanghai, China}

\begin{abstract}
We present the experimental generation of large effective magnetic fields for ultracold atoms using photon-assisted tunneling in an optical superlattice. The underlying method does not rely on the internal structure of the atoms and therefore constitutes a general approach to realize widely tunable artificial gauge fields without the drawbacks of near-resonant optical potentials. When hopping in the lattice, the accumulated phase shift by an atom is equivalent to the Aharonov-Bohm phase of a charged particle exposed to a staggered magnetic field of large magnitude, on the order of one flux quantum per plaquette. We study the ground state of this system and observe that the frustration induced by the magnetic field can lead to a degenerate ground state for non-interacting particles. We provide a local measurement of the phase acquired by single particles due to photon-assisted tunneling. Furthermore, the quantum cyclotron orbit of single atoms in the lattice exposed to the effective magnetic field is directly revealed. 

\end{abstract}

\maketitle
\section{Introduction}
\label{intro}
The application of strong magnetic fields to two-dimen\-sional electron gases has led to the discovery of seminal quantum many-body phenomena, such as the integer and fractional quantum Hall effect \cite{tsui1982two}. Ultracold atoms constitute a unique experimental system for studying such systems in a clean and well controlled environment and for exploring new physical regimes, not attainable in typical condensed matter systems \cite{bloch2008many,fetter2009rotating}. However, charge neutrality of atoms prevents direct application of the Lorentz force with a magnetic field. An equivalent effect can be provided by the Coriolis force in a rotating atomic gas, which leads to the formation of quantized vortices in a Bose-Einstein condensate \cite{madison2000vortex}. The regime of fast rotation, in which the atomic gas occupies the lowest Landau level, was achieved in Refs.~\cite{schweikhard2004rapidly} but the amplitude of the effective gauge field remained too small to enter the strongly correlated regime that requires a number of vortices on the order of the particle number \cite{bloch2008many,cooper2001quantum}. An alternative route consists in applying Raman lasers to the gas in order to realize a Berry's phase for a moving particle \cite{lin2009synthetic,dalibard2010artificial}. The effective gauge fields generated in such a setup resulted in the observation of a few vortices, but were still far from the strong-field regime. 

Ultracold atoms in optical lattice potentials seem to constitute a promising system to enter the high-field regime. This has been shown in several theoretical proposals \cite{jaksch2003creation,gerbier2010gauge,mueller2004artificial}, which are based on the coupling between the motional degree of freedom and the internal structure of the atoms. The main challenge in lattice systems is to engineer complex tunneling amplitudes, that are spatially dependent, $K=|K| e^{i\phi(\mathbf{r})}$. This representation of the effect of a vector potential is known as the Peierls substitution, where $\phi(\mathbf{r})$ is the Peierls phase \cite{peierls1933energy,hofstadter1976energy}. Recently, tunable Peierls phases have been realized experimentally in an effective 1D ``Zeeman lattice'' using a combination of radio-frequency and Raman coupling \cite{jimenezGarcia2012peierls} and in periodically driven one-dimensional lattices \cite{Struck2012tunable}. 

In this Letter, we describe the creation of effective staggered magnetic fields for ultracold atoms in a two-dimensional optical lattice. The method we use is closely related to the proposal by Jaksch and Zoller \cite{jaksch2003creation} and subsequent work \cite{gerbier2010gauge}. However, it does not rely on the internal structure of the atom. We make use of the well known phenomenon of photon-assisted tunneling \cite{Eckardt2005localization,Lignier2007,Chen2011controlling}, as proposed by Kolovsky \cite{kolovsky2011creating}, using a pair of far-detuned run\-ning-wave laser beams. This allows us to engineer spa\-tial\-ly-dependent Peierls phases $\phi(\mathbf{r})$. The resulting effective magnetic flux through one unit cell of the lattice is fully tunable by the wavelength of the running-wave beams and their geometry. In our setup, the magnetic flux per four-site plaquette is staggered with a zero mean, alternating between $\pi/2$ and $-\pi/2$ \cite{lim2008staggered}. We study the nature of the ground state in this optical lattice by measuring its momentum distribution and show in particular that the frustration associated with the effective magnetic field can lead to a degenerate ground state for single particles, similar to the prediction of Ref.~\cite{moeller2010condensed}. We also study the quantum cyclotron dynamics of single atoms restricted to a four-site plaquette. Te main results we present here are published in Ref.~\cite{aidelsburger2011experimental}. The purpose of this paper is to provide a more detailed discussion of the underlying method and to present several techniques, that are relevant for the experimental realization. A similar scheme was proposed by Ref.~\cite{Bermudez2011Synthetic} to create Abelian and non-Abelian gauge fields with trapped ions, that can be also applied to other systems such as ultracold atoms in optical lattices.

\section{General idea}
\label{sec:1}
In this section we briefly discuss the general idea of the method used in our experiment. Consider atoms in a 2D lattice that is tilted along one direction and subjected to a periodic on-site modulation with site-dependent phases $\phi_{m,n}$. In the tight-binding approximation the dynamics of this system can be described by the following Hamiltonian:

\begin{eqnarray}
\hat{H}(t)&=& -J_x \sum \limits_{m,n} \left(\hat{a}_{m+1,n}^{\dagger} \hat{a}_{m,n} + \textrm{h.c.}\right) \nonumber\\
&\ & -J_y \sum \limits_{m,n} \left(\hat{a}_{m,n+1}^{\dagger} \hat{a}_{m,n} + \textrm{h.c.}\right) \nonumber\\
&\ & + \sum \limits_{m,n} \left[m \Delta + V_K^0 \cos(\omega t +\phi_{m,n}) \right] \hat{n}_{m,n}\ ,
\label{eq:drivingham}
\end{eqnarray} 

\noindent where $\hat{a}_{m,n}$ ($\hat{a}_{m,n}^{\dagger}$) annihilates (creates) a particle on site $(m,n)$ and $\hat{n}_{m,n}$ is the corresponding number operator. The strength of the tunnel coupling between neighboring lattice sites in the $x(y)$-direction is given by $J_{x(y)}$. The magnitude of the linear potential is given by $\Delta$, while the on-site modulation amplitude is $V_K^0$. For $\Delta \gg J_x$ tunneling is inhibited along the x-direction and can be restored by modulating the lattice on resonance $\hbar \omega=\Delta$. Using the Floquet formalism \cite{Grossmann1992localization,Holthaus1992collapse} the system can be described by an effective time-independent Hamiltonian

\begin{eqnarray}
\hat{H}_{eff}&=& -K \sum \limits_{m,n} \left(\hat{a}_{m+1,n}^{\dagger} \hat{a}_{m,n}\ e^{i\phi_{m,n}}+ \textrm{h.c.}\right) \nonumber\\
&\ & -J \sum \limits_{m,n} \left(\hat{a}_{m,n+1}^{\dagger} \hat{a}_{m,n} + \textrm{h.c.}\right) ,
\label{eq:effdrivingham}
\end{eqnarray} 

\noindent where $K$ is the effective tunnel coupling along $x$ and $J$ the effective tunnel coupling along $y$ determined by

\begin{eqnarray}
K&=& J_x\ \mathcal{J}_1\left(\frac{f(\phi_{m+1,n}-\phi_{m,n},V_K^0)}{\Delta} \right) \nonumber\\
J&=& J_y\ \mathcal{J}_0\left(\frac{f(\phi_{m,n+1}-\phi_{m,n},V_K^0)}{\Delta} \right) .
\label{eq:effcoupl}
\end{eqnarray} 

\noindent Here $\mathcal{J}_{\nu}(x)$ are the Bessel-functions of the first kind and $f(\eta,\xi)$ determines the differential modulation amplitude between neighboring sites \cite{hauke2012nonabelian}. For a vanishing phase difference $\eta=0$, also the differential modulation amplitude and therefore the induced coupling vanishes. Following the Floquet analysis, coupling is induced along the $x$-direction proportional to the first order Bessel-function $\mathcal{J}_{1}(x)$, while tunneling in the perpendicular direction is modulated by the zeroth order Bessel-function $\mathcal{J}_{0}(x)$ \cite{Eckardt2005localization,Creffield2013creating}. The induced phase factor $e^{i\phi_{m,n}}$ in Hamiltonian (\ref{eq:effdrivingham}) can give rise to a non-vanishing phase $\Phi=\phi_{m,n+1}-\phi_{m,n}$ for an atom that tunnels along the borders of one plaquette. This can be interpreted as an Aharonov-Bohm phase and therefore serves as a model system to study artificial magnetic fields.

\begin{figure}
\resizebox{\columnwidth}{!}{%
  \includegraphics{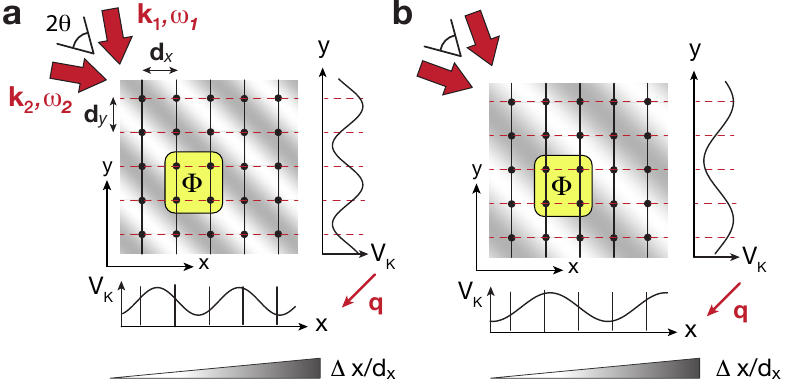}
	}
\caption{Lattice setup to create tunable artificial magnetic fields. It consists of a 2D array plus a linear potential of strength $\Delta$ along the $x$-direction. A pair of running-wave beams with wave vectors $|\mathbf{k}_1|\simeq|\mathbf{k}_2|=k$ and a frequency difference $\omega_1-\omega_2=\omega$ is used to modulate the lattice on resonance, so that the corresponding optical potential $V_K$ gives rise to a spatially dependent complex coupling. Here a snapshot of the running-wave potential $V_K$ with wave vector $\mathbf{q}$ is shown. The resulting effective magnetic flux $\Phi$ per plaquette (highlighted by the yellow box) is fully determined by the opening angle $2\theta$ and the frequency of the running-wave beams. Two examples are illustrated here for $k=\pi/d_y=\pi/d_x$ and \textbf{(a)} $\theta=35^{\circ}$ and \textbf{(b)}  $\theta=25^{\circ}$.}
\label{Fig_Method}   
\end{figure}

Experimentally, Hamiltonian (\ref{eq:effdrivingham}) can be realized using two running-wave beams with wave-vectors $\mathbf{k}_{1,2}$ ($|\mathbf{k}_1|\simeq|\mathbf{k}_2|=k$) and frequencies $\omega_{1,2}$ such that $\omega_1 - \omega_2 =\omega$. The local optical potential created by these two running-wave laser beams is given by

\begin{equation}
V_K(\mathbf{r}) = V_K^0 \cos^2(\mathbf{q} \mathbf{r} /2-\omega t /2) ,
\label{eq:interference}
\end{equation}

\noindent where $\mathbf{q}=(\mathbf{k_1}-\mathbf{k_2})$, $|\mathbf{q}|=2 k \sin \theta$ and $\theta$ is half of the opening angle between the two beams. This potential gives rise to time-dependent on-site modulation terms with spatial dependent phases given by the following relation:
\begin{equation}
\phi_{m,n}= \alpha_x m + \alpha_y n , \ \ \textrm{with}\ \alpha_i=q_i\cdot d_i ,
\end{equation}

\noindent $d_i$ being the lattice constant along the corresponding axis and an effective flux $\Phi=\alpha_y$. Note that even though the photon-assisted tunnel coupling occurs along the $x$-direction, the flux $\Phi$ through one plaquette is solely determined by the projection of the wave vector $\mathbf{q}$ on the perpendicular $y$ axis. Its value is fully tunable through the frequency and the angle between the pair of running-wave beams. Figure \ref{Fig_Method} illustrates the tunability of the flux with the angle $\theta$. In Fig.~\ref{Fig_Method}(a) two beams with an opening angle of $\theta=35^{\circ}$ and a flux of $\Phi\approx0.8\pi$ are shown, while in Fig.~\ref{Fig_Method}(b) the angle is slightly different, $\theta=25^{\circ}$, realizing a flux of $\Phi \approx0.6\pi$. 

In the original proposal by Kolovsky \cite{kolovsky2011creating} the photon-assisted tunneling along the $x$-direction was implemented by displacing the crossing point of the running-wave beams with respect to the atom cloud position. However, tunneling along the perpendicular direction is also affected by the lattice modulation, such that this scheme leads to non-uniform tunnel couplings $J$ and therefore this scheme cannot be employed to simulate uniform magnetic fields as shown recently by Creffield and Sols \cite{Creffield2013creating}. 

\section{Experimental Setup}
\label{sec:2}
\begin{figure}
\resizebox{\columnwidth}{!}{%
  \includegraphics{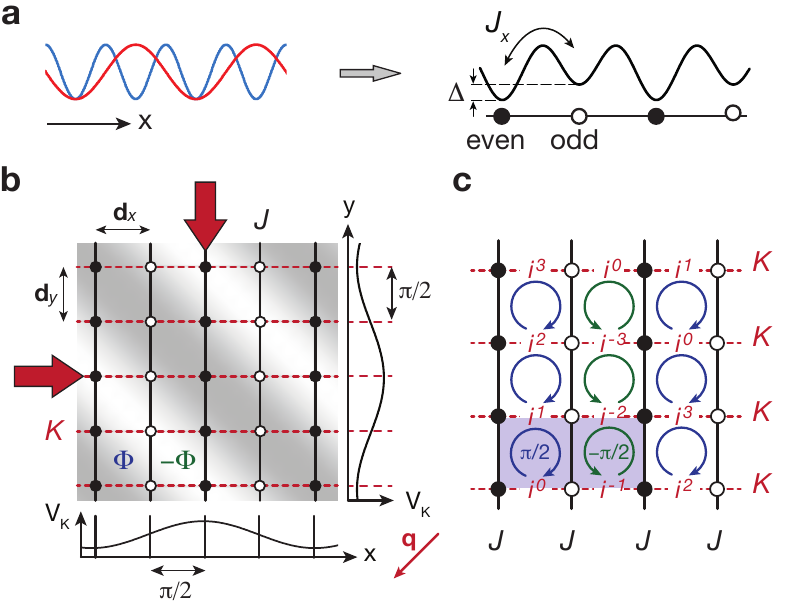}
	}
\caption{Experimental setup. \textbf{(a)} Superposition of two standing waves at $\lambda_s=767$ nm (blue) and $\lambda_l=1534$ nm (red) leading to a staggered energy offset. Black filled circles illustrate sites of low energy (even) and white circles the ones of high energy (odd). \textbf{(b)} The experimental setup consists of a 2D array of 1D potential tubes with spacing $|\mathbf{d}_x|=|\mathbf{d}_y|\equiv d=\lambda_s/2$. 
While bare tunneling occurs along the $y$ direction with amplitude $J$, it is inhibited along $x$ due to the superlattice potential. A pair of far-detuned running-wave beams creates an additional running-wave potential $V_K$ with wave vector $\mathbf{q}$, which induces a complex tunnel coupling of magnitude $K$, whose phase depends on position. This realizes an effective flux $\pm \Phi$ per plaquette with alternating sign along $x$. \textbf{(c)} Spatial distribution of the complex phase of the effective tunnel coupling realized in the experiment. The shaded area highlights the magnetic unit cell. Figure adapted from Ref.~\cite{aidelsburger2011experimental}.}
\label{Fig_Scheme}       % Give a unique label
\end{figure}

The experimental setup consists of a 2D lattice created by two standing waves of laser light at $\lambda_s=767$~nm (``short lattice'', see also Ref.~\cite{aidelsburger2011experimental}). Instead of a linear potential in the $x$-direction we superimpose another standing wave along that axis with twice the wavelength (``long lattice'', $\lambda_l \approx 2 \lambda_s$) to create a superlattice potential \cite{sebbystrabley2006superlattice,foelling2007direct}, where every second lattice site is lifted by an energy offset $\Delta$ as depicted in Fig.~\ref{Fig_Scheme}(a). 

The lattice parameters are chosen such that $\Delta$ is much larger than the nearest-neighbor coupling $J_x$ in order to inhibit tunneling. By resonantly modulating the lattice with two running-wave beams with wavelength $\lambda_l\simeq\,1534$\,nm ($\hbar\omega=\Delta$) tunneling is restored with an effective coupling strength $K$. To avoid cross-interference between the long lattice and the pair of running-wave laser beams we offset them in frequency by $\simeq\,160\,$MHz. As explained in the previous section, the modulation by the running-wave potential leads to a complex phase factor that depends on the position in the lattice and is determined by the geometry and the wavelength of the running-wave beams. In this setup the wave vectors are chosen to be $|\mathbf{k}_1|\simeq|\mathbf{k}_2|=k_s/2$, with $k_s=2\pi/\lambda_s$, and the angle between the beams is $2\theta=90^{\circ}$ as illustrated in Fig.~\ref{Fig_Scheme}(b), so that the optical potential is given by

\begin{equation}
V_K(\mathbf{r}) = V_K^0 \cos^2\left( \frac{k_s}{4} (\mathbf{e}_x+\mathbf{e}_y)\mathbf{r}-(\omega_1-\omega_2) t /2\right) 
\label{eq:runningpotential}
\end{equation} 

\noindent and $q_y=k_s/2$.

Let us consider an atom tunneling from a site of low energy (even sites) to a site of high energy (odd sites). Assuming that $\omega_1>\omega_2$, the phase factor is given by $\phi_{m,n}=\frac{\pi}{2}(m+n)$ as illustrated in Fig.~\ref{Fig_Scheme}(b) and $\Phi=\pi /2$. For atoms that tunnel from odd to even sites, the sign of the energy offset an therefore the sign of the phase changes: $\phi_{m,n}=-\frac{\pi}{2}(m+n)$. The resulting spatial phase pattern for our system is shown in Fig.~\ref{Fig_Scheme}(c). Due to the staggering we realize an effective magnetic flux that is alternating in sign along the direction of the superlattice potential and the system can be described by the effective Hamiltonian

\begin{eqnarray}
\hat{H}&=& -K \sum \limits_{m,n} \left(\hat{a}_{m+1,n}^{\dagger} \hat{a}_{m,n} \ e^{\pm i\phi_{m,n}} + \textrm{h.c.}\right) \nonumber\\
&\ & -J \sum \limits_{m,n} \left(\hat{a}_{m,n+1}^{\dagger} \hat{a}_{m,n} + \textrm{h.c.}\right) ,
\label{eq:mainham}
\end{eqnarray} 

\noindent The sign of the phase factor is positive for even sites and negative for odd ones. In the limit $\Delta > V_K^0$ the effective couplings are given by the following equations:

\begin{eqnarray}
K&=& J_x \mathcal{J}_1(V_K^0/(\sqrt{2}\Delta)) \approx J_x V_K^0/(2\sqrt{2}\Delta)\nonumber\\
J&=& J_y \mathcal{J}_0(V_K^0/(\sqrt{2}\Delta)) \approx J_y .
\label{eq:effcoupl}
\end{eqnarray} 

\noindent with $f(\pi/2,V_K^0)=V_K^0/(\sqrt{2})$, see Sect.~\ref{sec:32} for more details.

The magnetic unit cell of a staggered flux lattice consists of two plaquettes with opposite sign of the flux as illustrated by the shaded area in Fig.~\ref{Fig_Scheme}(c), see also Ref.~\cite{moeller2010condensed}. It contains two non-equivalent sites (even and odd), therefore the band splits into two sub-bands \cite{blount1962bloch,wang2006hofstadter}. Considering the magnetic translation symmetries we can make the following ansatz for the wave function using the gauge realized in our setup:

\begin{equation}
\psi = \sum \limits_{m,n} \textrm{e}^{\ii (m\cdot k_x d + n\cdot k_y d)} \times  \left\{ {
   \begin{array}{ll}
    \psi_e &  \quad m\ \textrm{even}\\
    \psi_o\ \textrm{e}^{\ii \frac{\pi}{2} (m + n)} & \quad  m \ \textrm{odd}\\
   \end{array} }\ , \right.
   \label{eq:wavefct}
\end{equation}

\noindent with the quasi-momentum $\mathbf{k}$ in the ranges $-\pi/(2d) \leq k_x < \pi / (2d)$ and $-\pi/d \leq k_y < \pi/ d$. By inserting this ansatz into the Schr\"odinger equation using the Hamiltonian given in Eq.~(\ref{eq:mainham}) one obtains a simple two-dimension\-al eigenvalue equation for the amplitudes $\psi_{e(o)}$ on even (odd) lattice sites:

\begin{equation}
\hat{H_0} \left( {
   \begin{array}{l}
    \psi_e\\
    \psi_o\\
   \end{array} } \right) = E \left( {
   \begin{array}{l}
    \psi_e\\
    \psi_o\\
   \end{array} } \right),
   \label{eq:eigenvalues}
\end{equation}

\noindent with

\begin{equation}
\hat{H_0} = \left( {
   \begin{array}{cc}
   -2J\ \textrm{cos}(k_y d) & -K(\mathrm{e}^{\ii k_x d} - \ii \textrm{e}^{-\ii k_x d}) \\
-K(\mathrm{e}^{-\ii k_x d} + \ii \mathrm{e}^{\ii k_x d}) & 2J\ \textrm{sin}(k_y d) \\
   \end{array} } \right). 
   \label{eq:simpleham}
\end{equation}

\noindent The solutions of this equation determine the dispersion relation of the system for any ratio $J/K$. Due to the intrinsic structure of the wave function given in Eq.(\ref{eq:wavefct}) every momentum eigenstate $\mathbf{k}$ has two momentum components at $(k_x,k_y)$ (even sites) and $(k_x,k_y)+(k_s/2,k_s/2)$ (odd sites). This means that for a non-degenerate ground state we expect to see two momentum peaks within the first magnetic Brillouin zone and replicas at momentum components shifted by the  reciprocal lattice vectors $(k_s,0)$ and $(0,2k_s)$.

\section{Calibration of the setup}
\label{sec:3}
In this section we give details about the underlying setup relevant for the experimental realization of the staggered flux lattice and calibration measurements to study the modulation induced coupling strength $K$.

\subsection{Staggered energy offset}
\label{sec:31}
\begin{figure}
\resizebox{\columnwidth}{!}{%
  \includegraphics{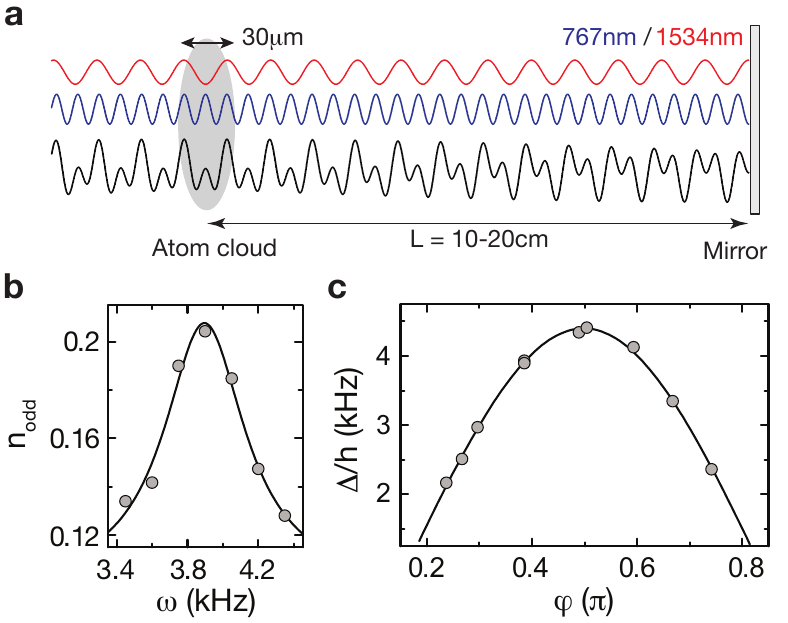}
	}
\caption{Calibration of the superlattice potential. \textbf{(a)} Illustration of the phase accumulation between the two standing waves, long (red) and short (blue), until they reach the atom position. Note that the distances shown here are not to scale.\textbf{(b)} Fraction of atoms transferred to odd lattice sites $n_{\rm{odd}}$ as a function of the modulation frequency $\omega=\omega_1-\omega_2$ for $\varphi=0.39\pi$. The solid line shows the fit of a Lorentzian function to our data.
\textbf{(c)} Energy offset $\Delta$ between even and odd lattice sites measured as a function of the relative phase $\varphi$ using photon-assisted tunneling. The maximum offset corresponds to $\varphi=\pi/2$, which is the staggered configuration. The solid line is a guide to the eye.}
\label{Fig_phase}       % Give a unique label
\end{figure}

It is important for the implementation of the staggered flux lattice that the coupling between even and odd sites is equal and does not depend on the lattice site. An example of such a potential is shown in Fig.~\ref{Fig_calib}(a). The particular shape of the superlattice structure depends on the relative phase $\varphi$ between the short- and long-lattice standing waves and their depths, which can be controlled independently. The combined potential is given by the following equation: 

\begin{equation}
V(x)=V_l^x \sin^2(k_l x+\varphi/2)+V_s^x \sin^2(2k_lx + \pi/2),
\end{equation}

\noindent where $k_l=2\pi/\lambda_l$ and $V_{l(s)}^x$ is the depth of the long (short) lattice. The phase of both standing waves is fixed at the position of the retro-reflecting mirror. Due to a small frequency difference between the two standing waves $\delta\nu=2\nu_l-\nu_s$, where $\nu_{l(s)}$ is the frequency of the long (short) lattice, they accumulate a relative phase $\varphi \propto \delta\nu L$ that depends on the distance $L$ between the retro-reflecting mirror and the atoms, see Fig.~\ref{Fig_phase}(a). Over the extent of the atom cloud the phase can be assumed constant to very good approximation. For our experimental setup it changes by $<1\rm{mrad}$ over the cloud size. By changing the frequency of one of the beams we have full control over the superlattice potential. 

In order to calibrate the relative phase $\varphi$, we measured the energy offset $\Delta$ between neighboring sites using photon-assisted tunneling. The experimental se\-quence started by adiabatically loading a condensate of $^{87}$Rb atoms without discernible thermal fraction into the 2D lattice for a given value of $\varphi$, such that initially all atoms populated the wells of lower energy. The lattice parameters were  $V_l^x=5\,E_{rl}$, $V_s^x=9\,E_{rs}$ and $V_s^y=20\,E_{rs}$, where $E_{ri}=h^2/(2m\lambda_{i}^2)$, $i=l,s$, is the respective recoil energy. In addition we used a third lattice along the vertical direction, with $\lambda_z=844\,$nm and $V^z=20\,E_{rz}$, where $E_{rz}=h^2/(2m \lambda_z^2)$, to avoid coupling to longitudinal modes. Then we switched on the running-wave beams for a finite time, varying their frequency difference $\omega$ for different experimental runs. When $\omega=\Delta/\hbar$ we observed resonant particle transfer between even and odd lattice sites. Figure~\ref{Fig_phase}(c) shows the results of the offset calibration as a function of the relative phase. The highest energy difference occurs for $\varphi=\pi/2$ and corresponds to the staggered potential, where the value of the offset is $\Delta/\rm{h}=4.4(1)$\,kHz. An exemplary resonance curve is shown in Fig.~\ref{Fig_phase}(b), where we measured the fraction of atoms on odd sites $n_{\rm{odd}}=N_{\rm{odd}}/(N_{\rm{even}}+N_{\rm{odd}})$ as a function of the modulation frequency, where $N_{\rm{odd (even)}}$ is the atom number on odd (even) lattice sites. The modulation induced coupling strength $K/\rm{h}$ for the measurements presented in this section is on the order of $30\,$Hz, while the damping time of the tunnel oscillations due to inhomogeneities in our system is typically on the order of a few ms. The amplitude of the resonance is at maximum $0.5$, which corresponds to an equal population on even and odd lattice sites. To measure the atom population in even and odd sites we employed a detection sequence explained in Sect.~\ref{sec:32} (see also Fig.~\ref{Fig_detect}).

\subsection{Modulation induced tunnel coupling}
\label{sec:32}

 \begin{figure}
\resizebox{\columnwidth}{!}{%
  \includegraphics{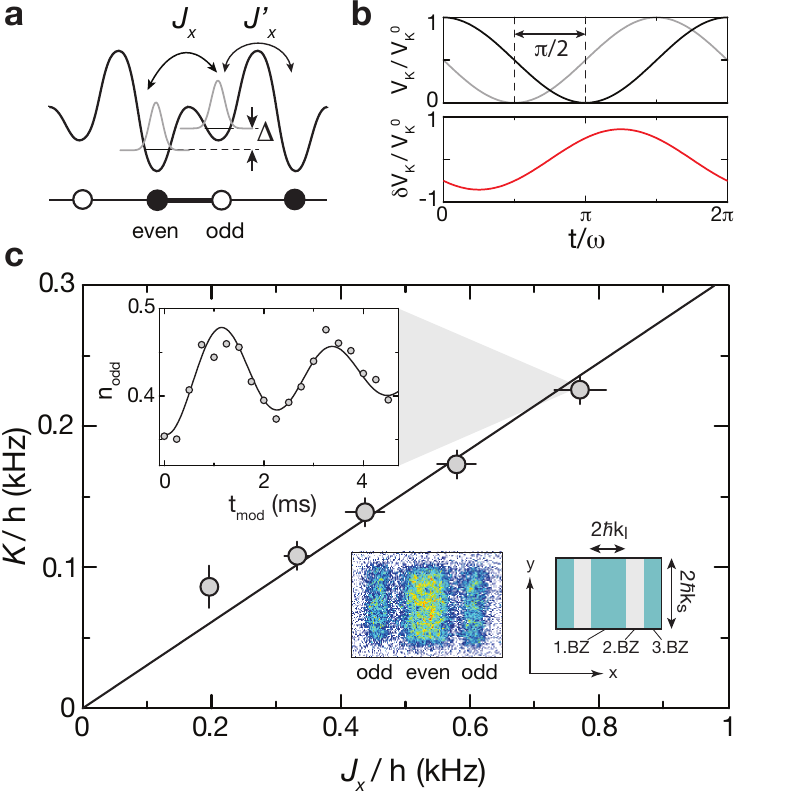}
	}
\caption{Calibration of the modulated tunnel coupling. \textbf{(a)} Schematics of a double well potential with energy offset $\Delta$, inner-well coupling $J_x$ and inter-well coupling $J'_x$. For these measurements $J_x$ was about two orders of magnitude bigger than $J'_x$.
\textbf{(b)} Modulation potential $V_K(t)$ as a function of time for even (black) and odd sites (gray). The differential modulation amplitude $f(\pi/2,V_K^0)$ is given by their difference and is shown in the lower graph (red curve).
\textbf{(c)} Measured coupling strength $K$/h as a function of the inner-well coupling $J_x$. The black curve is a linear fit to the data. The inset shows the tunnel oscillations between even and odd sites as a function of the modulation time $t_{mod}$ for $J_x/\rm{h}=0.77(4)$\,kHz. The black line shows the fit of a damped sine-wave and gives $K/\rm{h}=0.23(1)$\,kHz. The horizontal error bars show the uncertainty in the calibration of the lattice depths and the vertical ones show the errors of the fit of a damped sine-wave. A typical image after the band mapping sequence for site-resolved detection is shown in the lower right corner together with a schematic drawing of the Brillouin zones (BZ). Atoms that are located on even sites appear in the first BZ while atoms in odd sites are transferred to the third BZ.}
\label{Fig_calib}       % Give a unique label
\end{figure}

In order to calibrate the modulation induced coupling strength $K$, we performed measurements in isolated double wells, where the coupling between even and odd sites is much higher than the one between odd and even sites ($J_x\gg J'_x$), so that all dynamics were restricted to two sites only (see Fig.~\ref{Fig_calib}(a)). The corresponding parameter regime for the superlattice potential is $V^x_l \gg V^x_s$ and $0 \leq \varphi < \pi/4$. The magnitude of the modulated coupling strength is determined by the inner-well coupling $J_x$, the energy offset $\Delta$ and the differential modulation amplitude $f(\pi/2,V_K^0)$ (see also Sect.~\ref{sec:1}). Due to the form of the optical potential created by the pair of running-wave beams given in Eq.(\ref{eq:runningpotential}), there is a relative phase between the modulation of even and odd sites equal to $\pi/2$, so that

\begin{eqnarray}
V_{K,\rm{even}}(t)&=&\ V_K^0 \cos^2(\omega \cdot t/2) \ \ \rm{and}\nonumber\\
V_{K,\rm{odd}}(t)&=&\ V_K^0 \cos^2(\pi/4+\omega \cdot t/2) .
\end{eqnarray}

\noindent The differential modulation amplitude $f(\pi/2,V_K^0)$ is then given by their difference, which can be written as

\begin{equation}
V_{K,\rm{odd}}(t)-V_{K,\rm{even}}(t)= \frac{V_K^0}{\sqrt{2}} \sin({\omega \cdot t + \rm{cst}}) ,
\end{equation}

\noindent leading to $f(\pi/2,V_K^0)=V_K^0/\sqrt{2}$, see also Fig.~\ref{Fig_Scheme}(b). This leads to the following expression for the effective coupling \cite{Grossmann1992localization,Holthaus1992collapse}:

\begin{equation}
K= J_x \mathcal{J}_1\left(\frac{V_K^0}{\sqrt{2} \Delta} \right) \approx J_x V_K^0 /(2\sqrt{2} \Delta).
\label{eq:effcoupling}
\end{equation}

\noindent In order to verify this relation we performed a series of measurements to study the magnitude of $K$ as a function of $J_x$ for $\Delta$ and $V_K^0$ constant.

The experimental sequence started by adiabatically loading all atoms in a 3D lattice of depths $V_l^x=35\,E_{rl}$, $V^y_s=30\,E_{rs}$ and $V^z=30\,E_{rz}$ within $200\,$ms. Then we loaded all atoms adiabatically on the lower energy sites by ramping up the short lattice along $x$ in $10\,$ms. After switching on the pair of running-wave beams for a variable amount of time $t_{mod}$ we measured the fraction of atoms in odd sites $n_{\rm{odd}}$, see inset of Fig.~\ref{Fig_calib}(c). The evolution starts at $n_{\rm{odd}}=0$, because all atoms occupy the even sites of the system, then they start undergoing tunnel oscillations between even and odd sites, whose frequency is equal to $2K$/h. The strength of the inner-well coupling $J_x$ was controlled by the short lattice depth $V^x_s$, which was varied between $8.5$ and $13.5\,E_{rs}$. The corresponding values of the tunnel coupling $J_x$ were obtained from a band structure calculation. The energy offset $\Delta$ was chosen to be equal to $4.4$\,kHz. As it is shown in Fig.~\ref{Fig_calib}(c) we find a linear dependence of the coupling strength on the inner-well coupling with a slope of $K/J_x=0.31(1)$. From an independent calibration of the potential depth created by the two running-wave beams $V_K^0=4.2(1)\,E_{rl}$, we obtain a slope $V_K^0/(2\sqrt{2} \Delta)=0.34(2)$ and taking into account non-linear corrections due to the Bessel function we obtain a slope of $\mathcal{J}_1(V_K^0/$ $(\sqrt{2} \Delta))=0.32(1)$. For $J_x/\rm{h}=0.20(1)\,$kHz the damping time of the tunnel oscillations due to inhomogeneities in our system is on the order of the oscillation period. This explains the larger uncertainty for the measured modulation induced coupling strength $K/\rm{h}$ and most likely explains the larger deviation from the linear curve, see Fig.~\ref{Fig_calib}(c).

 \begin{figure}
\resizebox{\columnwidth}{!}{%
  \includegraphics{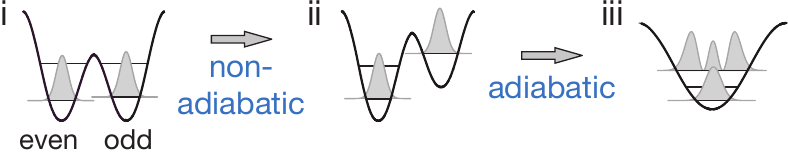}
	}
\caption{Site-resolved detection sequence. Atoms initially located in odd sites (i) are transferred non-adiabatically to the third energy level of the system (ii). After ramping down the short lattice adiabatically the occupation of the different Bloch bands is preserved (iii) and can be measured using a band-mapping technique.}
\label{Fig_detect}       % Give a unique label
\end{figure}

To detect the population on different sites we first ramped up the short lattice to freeze all tunneling dynamics and preserve the atom population on each lattice site. Then we transferred atoms that are located on odd sites non-adiabatically to the third energy level in the system by increasing the energy offset between neighboring sites. Ramping down the short lattice adiabatically preserved the population in the different Bloch bands, which were detected by a following band-mapping sequence (see also Fig.~\ref{Fig_detect}) \cite{sebbystrabley2006superlattice,foelling2007direct}. In the lower right corner of Fig.~\ref{Fig_calib}(c) we show a typical image obtained after the detection sequence together with the Brillouin zones. The size of the zones is reduced along the $x$-direction because the short lattice is ramped down earlier so that we are left with the long lattice only before the band-mapping sequence, which has a lattice constant that is twice the one of the short lattice.

\section{Ground state for isotropic coupling $J=K$}
The first measurement we present in this section was performed to study the ground state of the staggered flux lattice from its momentum distribution for isotropic coupling strengths $J=K$ \cite{aidelsburger2011experimental}. 

\begin{figure}
\resizebox{\columnwidth}{!}{%
  \includegraphics{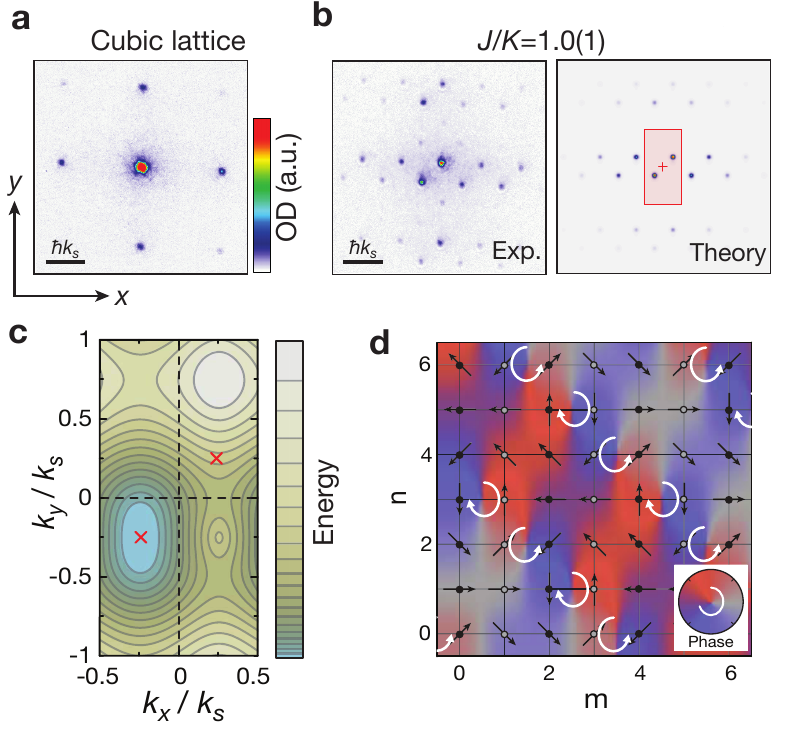}
	}
\caption{Ground state of the staggered flux lattice for isotropic coupling $J=K$.
\textbf{(a)} - \textbf{(b)} Momentum distribution measured after a time of flight of $t_{\rm{TOF}}=20$\,ms for (a) simple cubic lattice without modulated coupling by the running-wave beams and (b) staggered flux lattice as explained in Sect.~\ref{sec:2} for $J/K=1.0(1)$. The theoretical momentum distributions are obtained by an exact diagonalization of the Hamiltonian in Eq.~(\ref{eq:mainham}) on a $31 \times 31$ lattice with a harmonic confinement of 20~Hz. The red square indicates the magnetic Brillouin zone and the cross marks the center.
\textbf{(c)} Dispersion relation of the lowest Bloch band, calculated in the tight-binding approximation using the Hamiltonian given in Eq.~(\ref{eq:simpleham}) plotted for the first magnetic Brillouin zone. The red crosses mark the position of the ground-state momentum eigenstate together with the one shifted due to the intrinsic structure of the wave function given in Eq.(\ref{eq:wavefct}).
\textbf{(d)} Spatial distribution of the phase for the ground-state wave function. The vortices of different chirality are illustrated by the rotation of the white arrows. Figure adapted from Ref.~\cite{aidelsburger2011experimental}.}
\label{Fig_GSequal}       % Give a unique label
\end{figure}

The experimental sequence started by loading a condensate of about $5\times10^4$ atoms with no discernible thermal fraction into a 2D lattice within 160\,ms, forming an array of 1D Bose gases. The depth of the lattice was adjusted to $V^x_s=V^y_{s}=14\,E_{rs}$, which corresponds to an equal tunnel coupling along both directions of $J_x/{\rm{h}}=J/{\rm{h}}=31(2)$\,Hz. Then we ramped up the long-lattice in 0.7 ms to $V^x_l=5\,E_{rl}$ with $\varphi=\pi/2$ and subsequently decreased the short lattice in the $x$-direction to $9\,E_{rs}$ within $1$\,ms. The staggered energy offset between neighboring sites was $\Delta/{\rm{h}}= \rm{4.4}(1)$\,kHz and was calibrated using the measurements presented in Sect.~\ref{sec:31}. The bare tunnel coupling along $x$ was $J_x/{\rm{h}}=94(4)$\,Hz and therefore much smaller than $\Delta$. This results in an inhibition of spontaneous tunneling along this direction, which was then restored by switching on the running-wave beams on resonance, $\omega =2\pi \times 4.4$\,kHz and $V_K^0=4.2(1)\,E_{rl}$. This yields a value of $K/{\rm{h}}=30(2)$\,Hz and was obtained using the calibration measurements presented in Sect.~\ref{sec:32}. After holding the atoms in this configuration for $10$\,ms, we observed a restored phase coherence as shown in Fig.~\ref{Fig_GSequal}(b). Most likely, this can be attributed to a redistribution of entropy present in the random phases between the 1D condensates along their longitudinal modes.

In Fig.~\ref{Fig_GSequal}(a) we show the momentum distribution of a simple cubic lattice with lattice constant $d=\lambda_s/2$, without running-wave beams. In contrast to this simple structure with a single quasi-momentum component and correspondingly single momentum peaks separated by $2\hbar k_s$, we observe four momentum components within the first Brillouin zone of the short lattice in the presence of the staggered flux, see Fig.~\ref{Fig_GSequal}(b). In order to understand that distribution we calculated the dispersion relation using Eqs.~(\ref{eq:eigenvalues}) and (\ref{eq:simpleham}) for $J=K$, which is plotted in Fig.~\ref{Fig_GSequal}(c). It shows a single minimum located at $\mathbf{k}=(-k_s/4,-k_s/4)$ and the eigenvector has equal weight on even and odd sites, $|\psi_e|=|\psi_o|$. Therefore we expect to see two momentum components within the first magnetic Brillouin zone, one at $(-k_s/4,-k_s/4)$ (even sites) and a second one shifted due to the additional phase factor in the wave function for odd sites at $(k_s/4,k_s/4)$. The magnetic unit cell, illustrated by the shaded area in Fig.~\ref{Fig_Scheme}(c), contains two lattice sites, therefore the size of the magnetic Brillouin zone is reduced along the $x$-direction compared to the one of the square lattice. All other momentum peaks are translated by the reciprocal lattice vectors $(k_s,0)$ and $(0,2k_s)$. The measured positions are in good agreement with the quasimomenta of the Bloch states of lowest energy.

The frustration introduced by the position-dependent phase factors $\phi_{m,n}$ causes the phase of the atomic wave function to be non-uniform, see Fig.~\ref{Fig_GSequal}(d). The direction of the arrows on each lattice site indicates the phase of the wave function, while their length corresponds to the magnitude of the local atomic density. To make the appearance of vortices more clear we extrapolate the value of the phase between lattice sites and encode it in the background color of this plot. As one expects for an effective magnetic flux of $\Phi=\pi/2$, which is one quarter of a flux quantum, there is one vortex per four plaquettes along the $y$-direction and due to the staggering, the chirality of the vortices alternates along $x$. The atomic density of the ground-state wave function is uniform, illustrated by the length of the arrows in Fig.~\ref{Fig_GSequal}(d), as expected from the solution of the eigenvalue equation, that predicts equal weights for even and odd sites $|\psi_e|=|\psi_o|$.

\section{Ground state for anisotropic coupling}
\begin{figure}
\resizebox{\columnwidth}{!}{%
  \includegraphics{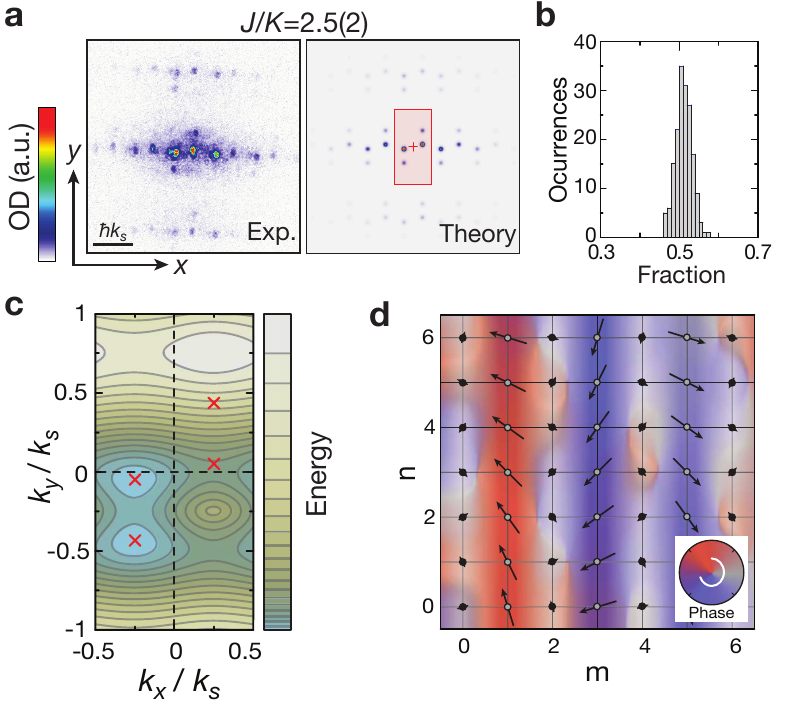}
	}
\caption{Ground state of the staggered flux lattice for $J=2.5K$.
\textbf{(a)} Momentum distribution measured after $t_{\rm{TOF}}=20$\,ms for $J/K=2.5(2)$. The theoretical momentum distributions are obtained by an exact diagonalization of the Hamiltonian in Eq.~(\ref{eq:mainham}) on a $31 \times 31$ lattice with a harmonic confinement of 20~Hz. The red square indicates the magnetic Brillouin zone and the cross marks the center.
\textbf{(b)} Histogram of the measured fraction of atoms in peaks corresponding to the lower momentum state, $\mathbf{k}_{\rm{L}}$ and $\mathbf{k}_{\rm{L}}+(k_s/2,k_s/2)$, obtained from 172 identical experimental runs.
\textbf{(c)} Dispersion relation of the lowest Bloch band, calculated using the Hamiltonian in Eq.~(\ref{eq:simpleham}) plotted for the first magnetic Brillouin zone. The red crosses mark the positions of the degenerate momentum ground state together with the ones shifted due to the intrinsic structure of the corresponding eigenstate.
\textbf{(d)} Spatial distribution of the ground-state phase distribution. The atomic density shows a charge density wave illustrated by the length of the arrows. A similar structure is obtained for the second degenerate momentum state but shifted by one lattice site. Figure adapted from Ref.~\cite{aidelsburger2011experimental}.}
\label{Fig_GSnotequal}       % Give a unique label
\end{figure}

To further study the nature of the ground state in our system we varied the ratio $J/K$ and observed the resulting momentum distribution of the ground state \cite{aidelsburger2011experimental}. The experimental sequence was similar to the one described in the previous section. In order to change the coupling ratio we kept the modulated coupling strength $K$ constant and changed the lattice depth along $y$ to $V^y_s=10\,E_{rs}$, which leads to $J/\rm{h}=75(3)$\,Hz and $J/K=2.5(2)$.

As can be seen in Fig.~\ref{Fig_GSnotequal}(a), we observed two additional peaks in the momentum distribution within the first magnetic Brillouin zone for these parameters. This is due to a two-fold degenerate ground state, in agreement with the solutions of the eigenvalue equation~(\ref{eq:eigenvalues}). The corresponding dispersion relation is plotted in Figure~\ref{Fig_GSnotequal}(c). It shows two minima at momenta $\mathbf{k}_{\rm{L}}\simeq(-0.25$ $k_s,-0.45$ $k_s)$ and $\mathbf{k}_{\rm{U}}\simeq(-0.25\,k_s,-0.052\,k_s)$, split a\-round $k_y=-k_s/4$ by $\Delta k_y \simeq 0.40\,k_s$. The relative atomic density on even and odd sites is given by the corresponding eigenvectors. For the eigenstate $\mathbf{k}_{\rm{L}}$ we obtain a relative weight $|\psi_o|^2/|\psi_e|^2\simeq 6.1$, which means that the momentum component shifted to $\mathbf{k}_{\rm{L}}+(k_s/2,k_s/2)\simeq(0.25\,k_s,0.052\,k_s)$ (odd sites) is about six times more intense than the one at $\mathbf{k}_{\rm{L}}\simeq(-0.25\,k_s,-0.45\,k_s)$ (even sites). For the second momentum eigenstate $\mathbf{k}_{\rm{U}}$ the relative weight is exactly opposite $|\psi_o|^2/|\psi_e|^2\simeq 1/6.1$ and the component at $\mathbf{k}_{\rm{U}}+(k_s/2,k_s/2)\simeq(0.25\,k_s,0.45\,k_s)$ (odd sites) is about six times less intense than the one at $\mathbf{k}_{\rm{U}}\simeq(-0.25\,k_s,-0.052\,k_s)$. This is in good agreement with our experimental observations, as shown in Fig.~\ref{Fig_GSnotequal}(a). The atomic density in the lattice is illustrated by the color brightness and the length of the arrows in Fig.~\ref{Fig_GSnotequal}(d) showing a charge density wave. The phase of the atomic wave function tends to align along the direction of the stronger coupling $J$, thereby frustrating the phase relation imposed by the Hamiltonian. As a consequence, the density in every second stripe along $y$ is suppressed. Figure~\ref{Fig_GSnotequal}(d) shows the result for the eigenstate $\mathbf{k}_{\rm{L}}$, where the odd sites are more populated. The result for $\mathbf{k}_{\rm{U}}$ is similar but shows a higher occupation for even sites.

\begin{figure}
\resizebox{\columnwidth}{!}{%
  \includegraphics{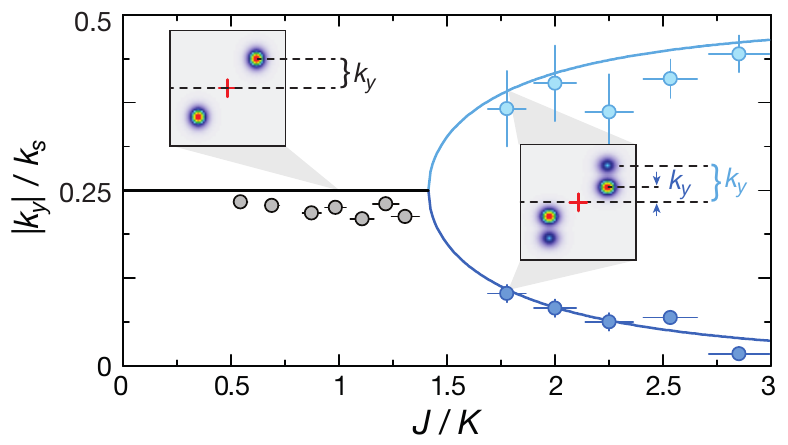}
	}
\caption{Projection of the momentum eigenstates (odd sites) on the $y$-direction as a function of the coupling ratio $J/K$. For $J/K<\sqrt{2}$ the peaks are located at $(k_s/4,k_s/4)$, while for $J/K>\sqrt{2}$ the peaks are split due to the emergent ground-state degeneracy, see insets. The solid lines correspond to the minima of the lowest Bloch band calculated in the tight-binding approximation using the Hamiltonian given in Eq.~(\ref{eq:simpleham}). Figure adapted from Ref.~\cite{aidelsburger2011experimental}.}
\label{Fig_bifurc}       % Give a unique label
\end{figure}

Contrary to the case of a triangular lattice with frustrated hopping, studied in Ref.~\cite{Struck21072011}, the fraction of atoms in each single-particle ground state does not fluctuate, as shown in Fig.~\ref{Fig_GSnotequal}(b). We observe equal population in both states as predicted for weakly interacting \cite{moeller2010condensed} or finite size systems. In trapped systems the degeneracy is slightly lifted and the two lowest energy states are symmetric and antisymmetric superpositions of the two momentum states.

We also varied the ratio $J/K$ gradually between $0.6$ and $2.8$ by adjusting the $y$-lattice depth to values between $9.5\,E_{rs}$ and $16.5\,E_{rs}$, while the modulated coupling strength $K/\rm{h}=30(2)$\,Hz was kept constant. The observed momentum distribution remained unchanged for $J/K \lesssim 1.4$, while above that value we observed the appearance of a two-fold degenerate ground state (see Fig.~\ref{Fig_bifurc}). For this measurement we focused on the momentum peaks that correspond to odd lattice sites ($k_x=k_s/4$) and plot their projection on the $y$-direction as a function of the ratio $J/K$. From the band-structure calculation we expect a bifurcation at $J/K=\sqrt{2}$ (solid line in Fig.~\ref{Fig_bifurc}), which agrees well with our data. In the limit of $J\gg K$ the two ground states approach $\mathbf{k}_{\rm{L}}\simeq(0.25\,k_s,0)$ and $\mathbf{k}_{\rm{U}}\simeq(0.25\,k_s,0.5\,k_s)$ with a splitting of $\Delta k_y=0.5\,k_s$, which corresponds to a state with relative atomic density $|\psi_o|^2/|\psi_e|^2\simeq 0$ and reversed for the other component. The nature of the bifurcation is identical to the one predicted in Ref.~\cite{moeller2010condensed}, where it is induced by a variation of the magnetic flux amplitude at a fixed value of $J/K=1$.

\begin{figure}
\resizebox{\columnwidth}{!}{%
  \includegraphics{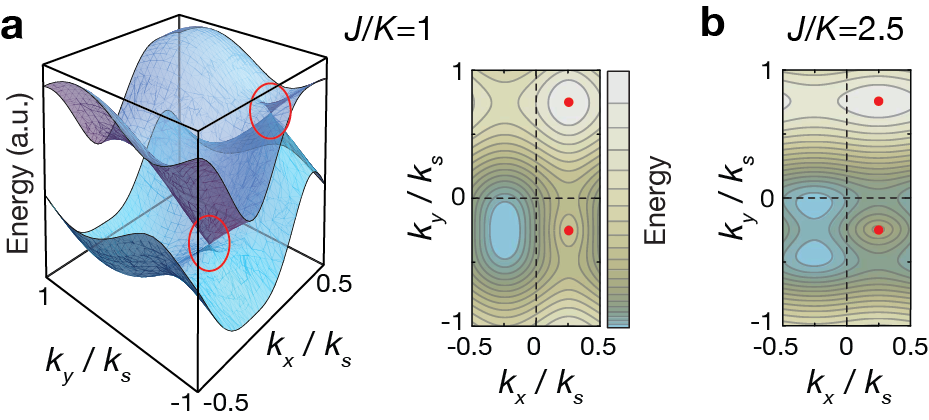}
	}
\caption{Band structure for \textbf{(a)} $J/K=1$ and \textbf{(b)} $J/K=2.5$. The positions of the Dirac cones are marked with red circles.}
\label{Fig_diraccones}       % Give a unique label
\end{figure}

To conclude this section we point out that there are two Dirac cones within the first Brillouin zone at momenta $\mathbf{k}_{\rm{D1}}\simeq(0.25\,k_s,-0.25\,k_s)$ and $\mathbf{k}_{\rm{D2}}\simeq(0.25\,k_s,0.75$ $k_s)$, see Fig.~\ref{Fig_diraccones}. It is interesting to mention, that their positions remain unchanged for different ratios of $J/K$ even in the regime, where there is a two-fold degenerate ground state whose momentum components are split around $k_y=-0.25\,k_s$, see Fig.~\ref{Fig_diraccones}(b).

\section{Local observation of the effective magnetic field}

\begin{figure}
\resizebox{\columnwidth}{!}{%
  \includegraphics{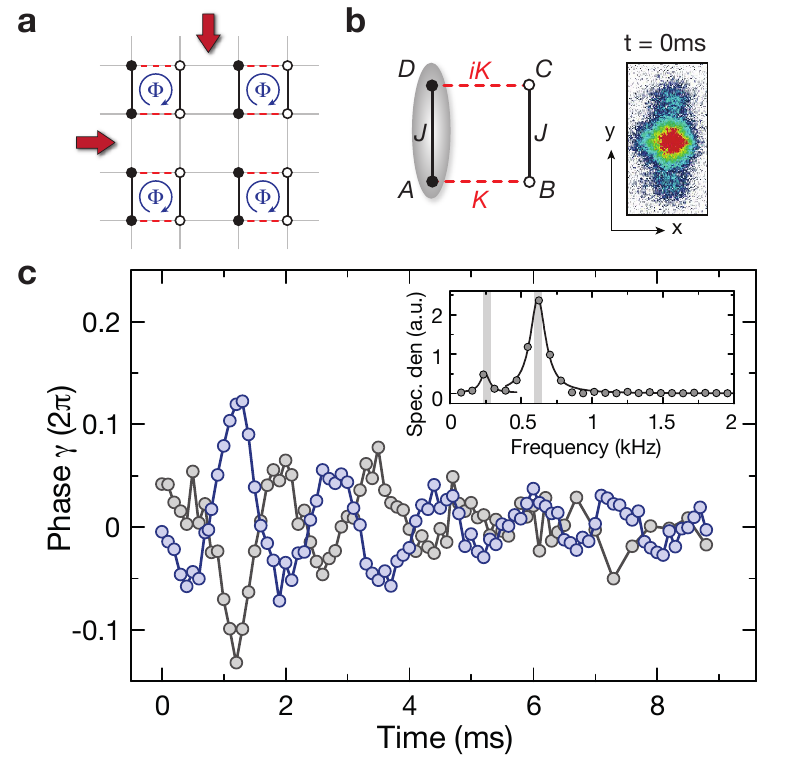}
	}
\caption{Local observation of the phase evolution in isolated plaquettes.
\textbf{(a)} Schematic illustration of the experimental setup. 
\textbf{(b)} The initial state is a single delocalized atom over the sites $A$ and $D$ (gray area), that is coupled to $B$ and $C$ sites using the pair of running-wave beams. On the right we show a typical double-slit pattern for the initial state obtained after $t_{\rm{TOF}}=20$\,ms.
\textbf{(c)} Phase evolution of the double-slit pattern along $y$ (integrated along $x$), as a function of time for $\hbar \omega=\Delta$ (blue) and $\hbar \omega=-\Delta$ (gray). The inset shows the Fourier transformation for $\hbar \omega=\Delta$ depicting two frequency components at $0.24(6)$\,kHz and $0.62(13)$\,kHz, in good agreement with theory (vertical lines). Figure adapted from Ref.~\cite{aidelsburger2011experimental}.}
\label{Fig_phasemeas}       % Give a unique label
\end{figure}

In order to probe the structure of the Hamiltonian in a more direct way, we performed a second series of experiments to study the local effect of the artificial gauge field on the level of a four-site square plaquette \cite{aidelsburger2011experimental}. The experimental setup is shown in Fig.~\ref{Fig_phasemeas}(a). We use two superlattice potentials along the $x$- and $y$-direction to isolate plaquettes so that all dynamics is restricted to four sites without any coupling between plaquettes, similar to the setup described in Sect.~\ref{sec:32}, where all dynamics was restricted to two sites only. This allows us to isolate plaquettes with equal sign of the flux. The relative phase between the two standing waves along $y$ was chosen to be $\varphi_y=0$, which corresponds to a symmetric double-well configuration without energy offset between neighboring sites. Along the $x$-direction the relative phase was in the regime $0 < \varphi_x < \pi/4$ in order to create a tilt between even and odd sites. To avoid coupling to axial modes along the potential tubes, an additional lattice along the $z$-direction was used. The four sites of a single plaquette are denoted as $A$, $B$, $C$ and $D$ (see Fig.~\ref{Fig_phasemeas}(b)).

The experimental sequence started by loading a condensate into a 3D lattice of depth $V^x_{l}=35\,E_{rl}$, $V^y_{l}=35\,E_{rl}$ and $V^z=30\,E_{rz}$. After that, a filtering sequence was applied to have at most one atom per lattice site \cite{foelling2007direct}. By ramping up the short-lattice along $x$ to $V^x_s=5\,E_{rs}$ within $10$\,ms and $\varphi=0.20(1)\,$rad we realized a double well potential with energy offset $\Delta/\rm{h}=6.0(1)$\,kHz, where single atoms were localized on one side. Finally the short-lattice along $y$ was ramped up to $V^y_s=14\,E_{rs}$ within $1$\,ms to create a plaquette potential, where each atom is in the initial state $\ket{\psi_1}=(\ket{A}+\ket{D})/\sqrt{2}$ ($J/\rm{h}=0.17(2)$\,kHz and $J_x/\rm{h}=2.0(1)$\,kHz). Subsequently the running-wave beams were switched on with $\omega =\Delta/\hbar$ in order to induce resonant coupling to the $B$ and $C$ sites. The induced coupling $K/\rm{h}=0.32(1)$\,kHz was measured independently.

The following rather complex phase evolution can be understood in the limit of $J\ll K$, where the dynamics along $y$ would be suppressed and the initial state $\ket{\psi_1}$ couples to the state $\ket{\psi_2}=(\ket{B}+i \ket{C})/\sqrt{2}$, due to the relative phase imprinted by the pair of running-wave beams. In our case the ratio is $J/K\approx 0.5$ leading to a more complex evolution. We measured the value of the phase as a function of time from the shape of the momentum distribution after time-of-flight. The method can be explained considering the double-slit interference obtained for an isolated double well along the $y$-direction, where we denote the two sites as $\ket{A}$ and $\ket{D}$, and we consider an arbitrary single-particle state

\begin{equation}
\ket{\psi}=\frac{1}{\sqrt{2}}(\ket{A}+\mathrm{e}^{i\gamma}\ket{D})
\end{equation} 

\noindent with phase $\gamma$. The observed density distribution after a sufficiently large $t_{\rm{TOF}}$ is proportional to  $\mathrm{cos}(k_t y+\gamma)$ times an envelope determined by the Wannier function, where $k_t=md/\hbar t_{\rm{TOF}}$ \cite{sebbystrabley2006superlattice,foelling2007direct}. For the initial state $\ket{\psi_1}$, $\gamma=0$ and we expect a symmetric momentum distribution. A typical image is shown in Fig.~\ref{Fig_phasemeas}(b). For the following evolution in isolated plaquettes, we first integrated the momentum distribution along $x$ and then determined the phase $\gamma(t)$ along $y$. The result is shown with blue data points in Fig.~\ref{Fig_phasemeas}(c). For $\omega=-\Delta/\hbar$ the role of the two running-wave beams is reversed, $\omega_1<\omega_2$ and the spatial phase pattern induced by the running-wave beams changes sign. Therefore we observed a sign reversal for the phase evolution.

\begin{figure}
\resizebox{\columnwidth}{!}{%
\includegraphics{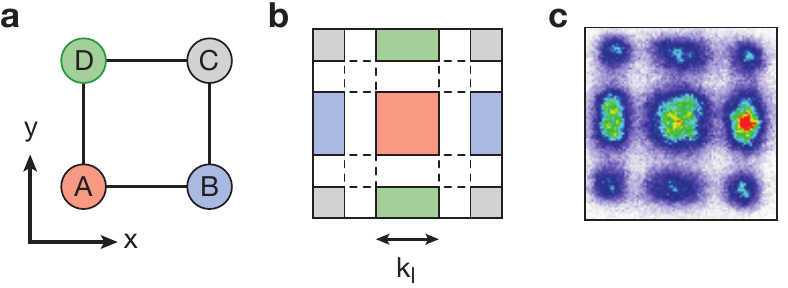}
	}
\caption{
\textbf{(a)} Schematic of the four-site plaquette.
\textbf{(b)} Brillouin zones of the 2D lattice.
\textbf{(c)} Typical momentum distribution obtained with the two-dimensional band mapping sequence after $20\,$ms of time-of-flight. Figure adapted from Ref.~\cite{aidelsburger2011experimental}.}
\label{fig:sitesplaq}
\end{figure}

We also observed the dynamics in real space in order to exhibit the influence of the artificial gauge field on the particle flow. The experimental sequence was the same than the one described above, only the depths of the short lattices were changed to $V^x_s=7\,E_{rs}$ and $V^y_s=10\,E_{rs}$ in order to obtain different coupling strengths $J_x/\rm{h}=1.2(1)\,$kHz and $J/\rm{h}=0.50(2)\,$kHz. By generalizing the site-resolved detection sequence discussed in Sect.~\ref{sec:32} for a system of isolated double wells to plaquettes, we measured the atom population on each site $N_q$ ($q=A, B, C, D$). First we applied the non-adiabatic mapping sequence along $x$ and $y$ in order to transfer the populations $N_q$ to different Bloch bands. Atom population originally in $B$ sites is therefore transferred to the third energy level in the $x$-direction, while population in $D$ sites is transferred to the third level along $y$. A subsequent band-mapping technique allowed us then to determine the occupation of different Bloch bands by counting the atom numbers in different Brillouin zones (see Fig.~\ref{fig:sitesplaq}(b)). The different colors show the connection between Brillouin zones and their corresponding lattice sites (see Fig.~\ref{fig:sitesplaq}(a)). A typical image obtained after $20\,$ms of time-of-flight is shown in Fig.~\ref{fig:sitesplaq}(c).

\begin{figure}
\resizebox{\columnwidth}{!}{%
  \includegraphics{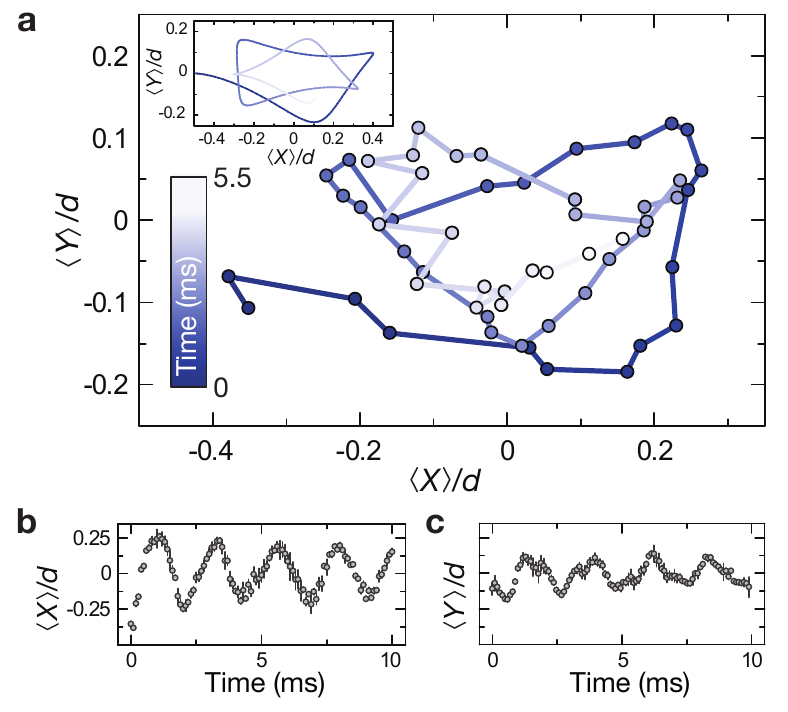}
	}
\caption{\textbf{(a)} Cyclotron orbits of the average particle position obtained from the mean atom positions $\left<X\right>/d$ \textbf{(b)} and $\left<Y\right>/d$ \textbf{(c)} for $J/\mbox{h}=0.50(2)\,$kHz, $K/\mbox{h}=0.28(1)\,$kHz and $\Delta/\mbox{h}=4.9(1)\,$kHz. Each data point is an average over three measurements. The inset in \textbf{(a)} shows the theoretical curve calculated for $\phi=0.80\times\pi/2$ and a $1/e-$damping time of $13\,$ms obtained from damped sine fits to $\left<X\right>/d$ and $\left<Y\right>/d$. Figure adapted from Ref.~\cite{aidelsburger2011experimental}.}
\label{Fig_current}       % Give a unique label
\end{figure}

From these images we extracted the average atom positions 

\begin{eqnarray}
\left<X\right>&=&\frac{(-N_A+N_B+N_C-N_D)\ d}{2N}  \\
\left<Y\right>&=&\frac{(-N_A-N_B+N_C+N_D)\ d}{2N}  \ ,
 \end{eqnarray}

\noindent with $N$ being the total atom number. Initially all atoms occupied the $A$ and $D$ sites with equal weight and the evolution starts at $\left<X\right>(t=0)=-0.5$ and $\left<Y\right>(t=0)=0$. After switching on the pair of running-wave beams we observed a coherent particle flow inside the plaquettes towards the $B$ and $C$ sites (see Fig.~\ref{Fig_current}(b)). Thereafter, the particle current showed deviations along the $y$-direction (see Fig.~\ref{Fig_current}(c)), which is  reminiscent of the Lorentz force acting on a charged particle in a magnetic field. As shown in Fig.~\ref{Fig_current}(a), the mean atom position followed an orbit that is a small-scale quantum analog of the classical cyclotron orbits for charged particles. This coherent evolution is damped due to spatial inhomogeneities in the atomic sample. Having independently calibrated the values of $J$ and $K$, we fit from the measured atom dynamics the value of the magnetic flux $\phi=0.73(5)\times\pi/2$. The difference from the value $\phi=\pi/2$ expected for a homogeneous lattice stems from the smaller distance between lattice sites inside the plaquettes when separated with an additional superlattice along the $y$-direction. For the parameters used in Fig.~\ref{Fig_current}(a)-(c) we calculate a distance $d_y=0.78(1)\times\lambda_s/2$ along $y$ yielding $\phi=0.80(1)\times\pi/2$, which qualitatively explains the measured flux value. Residual deviations might also be due to an angle mismatch between the two running-wave beams and the lattices beams. 

\section{Conclusion}
In conclusion, we have demonstrated a new type of optical lattice that realizes strong effective magnetic fields using a method that does not rely on the internal structure of the atom and is therefore applicable to a large variety of different systems. We have shown that the atomic sample relaxes to the minima of the magnetic bandstructure, realizing an analogue of a frustrated classical spin system. In addition we have presented a local measurement of the phase acquired by an atom subjected to the photon-assisted tunneling. Furthermore, the quantum cyclotron orbit of single atoms in the lattice has been observed and was used to evaluate the strength of the artificial magnetic field directly. The spatial average of the magnetic flux, however, is zero, hence the Bloch band is topologically trivial \cite{haldane1988model,wang2006hofstadter,gerbier2010gauge}. By using a superlattice potential with more than two non-equivalent sites \cite{gerbier2010gauge} or a linear tilt potential \cite{jaksch2003creation}, it is possible to create a lattice with a uniform and non-zero magnetic flux. This system would realize the Harper Hamiltonian \cite{harper1955single} and lead to the fractal band structure of the Hofstadter butterfly \cite{hofstadter1976energy}. In particular the lowest band would exhibit a Chern number of one and be analogous to the lowest Landau level \cite{jaksch2003creation,mueller2004artificial,dalibard2010artificial,cooper2011optical}. In addition the staggered flux lattice can be used to measure the $\pi$-flux associated with a Dirac cone using the concept of Zak phases following a recent proposal by D.~A.~Abanin \textit{et al.} \cite{Abanin2012interferometric}. This method was already demonstrated in the experiment for measuring topological invariants in 1D systems \cite{Atala2012direct}.

We acknowledge insightful discussions with N. Cooper and we thank J. T. Barreiro for careful reading of the manuscript. This work was supported by the DFG (FOR 635, FOR801), the EU (STREP, NAMEQUAM, Marie Curie Fellowship to S.N.), and DARPA (OLE program). M. Aidelsburger was additionally supported by the Deut\-sche Te\-lekom Stiftung.
% BibTeX users please use
% \bibliographystyle{}
% \bibliography{}
%
% Non-BibTeX users please use

\end{document}